\documentclass[letterpaper,12pt]{JHEP3}
\usepackage{amssymb}
\usepackage{mathrsfs}
\usepackage{cite}
\usepackage{texdraw}
\usepackage[english]{babel}

\csname @addtoreset\endcsname{equation}{section}

\newcommand{\C}{\mathbb{C}}
\newcommand{\R}{\mathbb{R}}

\newcommand{\Z}{\mathbb{Z}}

\newcommand{\A}{\mathcal{A}}

\newcommand{\LL}{\mathcal{L}}
\newcommand{\MM}{\mathcal{M}}
\newcommand{\NN}{\mathcal{N}}
\newcommand{\OO}{\mathcal{O}}
\newcommand{\RR}{\mathcal{R}}
\newcommand{\calS}{\mathcal{S}}

\newcommand{\dd}{\mathrm{d}}
\newcommand{\tr}{\mathrm{Tr}}
\newcommand{\erfc}{\mathrm{erfc}}
\newcommand{\p}{\partial}

\newcommand{\bra}[1]{\left< #1\right|}
\newcommand{\ket}[1]{\left| #1\right>}
\newcommand{\gym}{g_{\mathrm{YM}}}

\newcommand{\bea}{\begin{eqnarray}}
\newcommand{\eea}{\end{eqnarray}}

\def\al{\alpha}
\def\be{\beta}
\def\ga{\gamma}

\def\la{\lambda}

\def\si{\sigma}
\def\om{\omega}

\title{Giant gravitons in AdS/CFT (I):\\ matrix model and back reaction}
\author{Marco M. Caldarelli and Pedro J. Silva \\
Dipartimento di Fisica dell'Universit\`a di Milano, and\\
INFN, Sezione di Milano\\
Via Celoria 16, I-20133 Milano, Italy \\
E-mail: \email{marco.caldarelli@mi.infn.it},
\email{pedro.silva@mi.infn.it}}
\preprint{IFUM-797-FT \\
hep-th/0406096}

\abstract{
In this article we study giant gravitons in the framework of AdS/CFT
correspondence.
First, we show how to describe these configurations in the CFT side
using a matrix model. In this picture, giant gravitons are realized as
single excitations high above a Fermi sea, or as deep holes into
it. Then, we give a prescription to define quasi-classical states and
we recover the known classical solution associated to the CFT dual of
a giant graviton that grows in AdS.
Second, we use the AdS/CFT dictionary to obtain the supergravity
boundary stress tensor of a general state and to holographically
reconstruct the bulk metric, obtaining the back reaction of
space-time. We find that the space-time response to all the
supersymmetric giant graviton states is of the same form, producing
the singular BPS limit of the three charge Reissner-Nordstr\"om-AdS
black holes. While computing the boundary stress tensor, we comment on
the finite counterterm recently introduced by Liu and Sabra, and
connect it to a scheme-dependent conformal anomaly.
}


\begin{document}

\section{Introduction}
In the latest time the AdS/CFT conjecture has been one of the most studied 
subjects within string theory. This duality gives the possibility to study 
properties of supergravity theories from the CFT point of view and vice versa.
In particular it has brought new insights into black hole physics and the 
role of naked singularities in AdS (see
\cite{Petersen:1999zh,Aharony:1999ti,D'Hoker:2002aw} for reviews).

Although the initial studies in AdS/CFT where focused on the
supergravity approximation, lately the inclusion of different
D-brane configurations has enriched our understanding, exploring
new sectors that are necessary for the consistency of the whole
picture. In particular, since the AdS$_5$/CFT$_4$ duality was
constructed using D3-branes, stable D3-brane configurations in
AdS play an important role among all the different D-brane
sectors.

In this work we focus on a special type of D3-brane configurations called 
giant gravitons (GGs). These GGs are stabilized by a dynamical mechanism 
that develops local forces on the brane,
canceling its tension and avoiding therefore the worldvolume
collapse. Originally, GGs were described as D3-branes traveling on a
$\calS^1$ direction wrapping a perpendicular $\calS^3$, both contained
in the $\calS^5$ factor of the metric, while they sit on the center of
AdS$_5$ \cite{McGreevy:2000cw}\footnote{Nowadays we count with
  different generalization to this initial solution, where GGs travel
  along generic geodesics on AdS$_5$ \cite{Caldarelli:2004yk}, and
  where the D3-brane wraps general 3-cycles on the $\calS^5$
  \cite{Hackett-Jones:2004yi}. For a non-supersymmetric extension, see
  \cite{nonsusy}.}. The dynamics of the D3-brane effective
action allows for two different stable solutions, one in which the
radius of the $\calS^3$ is zero and the other with a non-vanishing
radius, bounded from above and proportional to the momentum along the
$\calS^1$ direction. Short after these configurations were studied,
another solution was found, where this time the D3-brane wraps an
$\calS^3$ inside the AdS$_5$ part of the ten-dimensional space-time
\cite{Hashimoto:2000zp}. All these three solutions share the same
quantum numbers and charges. In particular, they preserve half of the
supersymmetries \cite{Grisaru:2000zn} and, from the ten-dimensional
point of view, their geometrical center travels along a null
geodesic.

In this article, the first type of configuration will be called
{\it GG in $\calS^5$} while the second kind will be called {\it GG in
  AdS$_5$}. The point-like configuration corresponds to a
degenerate solution of the Born-Infeld action and will not be
considered here.
Originally, these configurations where thought as the gravitational
manifestation of the stringy exclusion principle
\cite{Maldacena:1998bw}, where the upper bound on the giant graviton
momentum on the $\calS^1$ (due to the fact that it is proportional to
the radius of the $\calS^3$ and therefore has a maximum on $\calS^5$),
is dual to the upper bound found on the conformal weight of a family of
chiral operators (here, the bound is easily understood from the finite
rank of the gauge symmetry group $U(N)$) \cite{McGreevy:2000cw,Balasubramanian:2001nh}.
But, unfortunately, the second kind of configuration, the GG in AdS,
has a completely different behavior, with no such upper bound.
This fact, added to the existence of the singular solution, shows
a not so clear picture (see \cite{Bena:2004qv} for a discussion of
this point).

GGs have also been studied from the dual CFT point of view. In
\cite{Balasubramanian:2001nh}, Balasubramanian et. al. proposed a
particular class of operators called sub-determinants as the duals of
GGs in $\calS^5$. Later on, Corley et. al. \cite{Corley:2001zk} extended this
proposition to describe all GGs in terms of a particular combination of
single trace and multi-trace operators on the real scalars of the CFT
theory, known as Schur polynomials. On the top of the above picture, GGs
in AdS also have a dual description in terms of a semi-classical
solution of the CFT theory \cite{Hashimoto:2000zp}, that remains to be
connected to the quantum theory (such links are conjectured in
\cite{Corley:2001zk,Berenstein:2004kk}). These semi-classical states
provide a picture where GGs may be seen as single D3-branes separating
from the initial stack of $N$ D3-branes.

{\em In this work, our first purpose is to define a theoretical framework
simple enough to be manageable, but containing the principal
physical ingredients that characterize these D3-brane states}.
To this end, we use a few basic assumptions to reduce the sector of
the CFT theory relevant for the discussion of GGs to
a quantum matrix model. Then, following the ideas of Hashimoto
et. al. \cite{Hashimoto:2000zp},
Corley et. al. \cite{Corley:2001zk} and Berestein
\cite{Berenstein:2004kk}, we arrive to a compact and elegant
description of the quantum system in terms fermionic degrees of freedom.
At this point, GG states are identified with
Schur polynomials acting on a Fermi sea. This procedure leads
to a beautiful interpretation of GGs as either holes in the
deep of the Fermi sea or as highly excited states
over the surface. Then, we identify quasi-classical states
and, as a particular case, recover the classical solution of
\cite{Hashimoto:2000zp}.

{\em Then, we proceed to our second purpose, which is to obtain the
  back reaction on the supergravity fields due to the presence of GGs
  in AdS$\times\calS^5$ space-time.}
To this end, we reconstruct the supergravity solution from the
boundary data using the AdS/CFT dictionary, and deduce that, at least
  asymptotically, the GG back reaction produces the singular BPS limit of the
Reissner-Nordstr\"om-AdS$_5$ family of solutions, with one or more
R-charges turned on \cite{Behrndt:1998ns}. In the quasi-classical
limit the branes localize in 
space-time and therefore the fields
they source are reproduced accurately by the supergravity solution
we obtain, up to regions of high curvature near the
singularity. This is where the GG condensate lies and supergravity
looses its validity.

These supergravity solutions were assigned originally to a GG
condensate in $\calS^5$ in \cite{superstars} by a different argument,
but from our analysis GGs in $\calS^5$ and GGs in AdS$_5$ produce the
same back reaction. This may seem surprising at first sight. In fact,
supergravity does not seem to be able to distinguish between different
GGs configurations with the same quantum numbers. Nevertheless, we
argue that the back reaction solution has different
ranges of validity for the two types of configurations and it makes
sense to trust the supergravity solution deep into the bulk only when
considering GGs in AdS.

To compare the supergravity and CFT stress energy tensors, we
need to compute the boundary stress tensor for the three-charge family
of Reissner-Nordstr\"om-AdS$_5$ solutions. In order to obtain the
results with the same renormalization scheme for the two sides of the
correspondence, we have to add precisely the finite counterterm
recently introduced by Liu and Sabra \cite{Liu:2004it} (see
\cite{odintsov} for previous works on the subject). Finally, we
argue that this boundary term is exactly the one needed to cancel the
scheme-dependent contribution to the conformal anomaly of the CFT.

The plan of the work is the following: in section~\ref{CFT}, we
introduce the CFT notations and conventions to define the dual
operators of GGs. Then, we describe the reduction of the CFT system
that leads us to a quantum mechanical matrix model and to the
description of GGs on it, including quasi-classical states. In
section~\ref{AdS}, we compute the expectation value of the quantum
stress energy tensor of the GG states, and match it with the
appropriate five dimensional supergravity solution. At last, in
section~\ref{END} we summarize and discuss our results. The details of
the calculations are left to the appendices. In
appendix~\ref{coherent} we give and justify the definition of coherent
and quasi-classical states for the reduced CFT model, while in
appendix~\ref{stresstensor} we perform the computation of the boundary stress
tensor and discuss the finite counterterm introduced by Liu and Sabra.

%%%%%%%%%%%%%%%%%%%%%%%%%%%%%%%%%%%%%%%%%%%%%%%%%%%%%%%%%%%%%%%%%%%%%%%%%%%%%%
%%%%%%%%%%%%%%%%%%%%%%%%%%%%%%%%%%%%%%%%%%%%%%%%%%%%%%%%%%%%%%%%%%%%%%%%%%%%%%

\section{Matrix model for giant gravitons}
\label{CFT}
Giant gravitons have been identified with a particular class of
half-BPS operators made out of the real scalars $X^m$ of $\NN=4$ SYM
theory\footnote{
In this article, we use greek indices $\mu,\nu,\ldots=0\ldots5$ for
the tangent space of five-dimensional space-time and latin indices
$a,b,\ldots=0,\ldots,3$ for the four-dimensional space-time on which
the CFT lives. The $U(N)$ and $SU(N)$ gauge group indices are
$i,j,\ldots=1,\ldots,N$ and for the R-charges we use $m,n,\ldots=1,\ldots,6$
indices when described in terms of the fundamental of $SO(6)$ and capital
$I,J,\ldots=1,2,3$ indices when described in terms of the fundamental
of $SU(4)$.
}.
We use in this article the $\NN=1$
decomposition, where the scalars are usually written as three complex
scalars $\Phi^I=\hbox{$1\over\sqrt 2$}(X^I+iX^{I+3})$, with $I=1,2,3$,
and all the fields transform in the adjoint representation of $U(N)$.
GGs are a combination of single-trace and multi-trace operators in
$\Phi^I$, labeled by their R-charge $n$. These operators are
then identified with Schur polynomial in $\Phi^I$, written either in
the totally symmetric representation $U$ of the associated symmetric
group $S_n$ (corresponding to a GG in AdS$_5$) or the totally
antisymmetric representation $U'$ of the symmetric group $S_n$
(corresponding to a GG in $\calS^5$) \cite{Corley:2001zk}.
To be more precise, Schur polynomial operators are defined as
\begin{equation}
  \chi_{(n,R)}\left(\Phi\right)={1\over n!} \sum_{\sigma \in S_n}
  \chi_R(\sigma)\,
    \Phi^{i_1}_{\sigma(i_1)}\cdots\Phi^{i_n}_{\sigma(i_n)}
\label{GG}\end{equation}
where, without loss of generality, we have set the SU(4) indices $I$ to $1$ and will neglect it for the rest of this section. We also have written explicitly the $U(N)$ indices `$i$',
taking values from $1$ to $N$. The sum is over all the group elements
$\sigma$ of the symmetric group $S_n$ and $\chi(\sigma)$ is the character
of the element $\sigma$ in the chosen representation $R$. The result of the
permutation $\si$ acting on the natural number `$i$' is
written as $\sigma(i)$. In particular, the $U$ and $U'$ representations
of $S_n$ have the following associated Young diagrams
$$
  \btexdraw
  \drawdim cm
%U
  \move(0 -.0)\lvec(3 -.0)\lvec(3 .5)\lvec(0 .5)\lvec(0 -.0)
  \move(0 0)\lvec(3 0)
  \move( .5 -.0)\lvec(.5  .5)
  \move(1   -.0)\lvec(1   .5)
  \move(1.5 -.0)\lvec(1.5 .5)
  \move(2.5 -.0)\lvec(2.5 .5)
  \move(1.8 .25)\fcir f:0 r:.01
  \move(2.0 .25)\fcir f:0 r:.01
  \move(2.2 .25)\fcir f:0 r:.01
  \htext(0.5 -1){$U$ symmetric}
%U'
 \move(5 .5)\lvec(5 -2.5)\lvec(5.5 -2.5)\lvec(5.5 .5)\lvec(5 .5)
 \move(5 .0)\lvec(5.5 0)
 \move(5 -.5)\lvec(5.5 -.5)
 \move(5 -2)\lvec(5.5 -2)
 \move(5.25 -.75)\fcir f:0 r:.01
 \move(5.25 -.95)\fcir f:0 r:.01
 \move(5.25 -1.15)\fcir f:0 r:.01
 \htext(6. -1){$U'$ antisymmetric}
 \etexdraw
$$
where the first is totally symmetric while the second is totally
antisymmetric\footnote{This diagrams are related to representations of
  $S_n$ and should not be confused with diagrams related to
  representations of $U(N)$.}. For example,
\begin{equation}
  \chi_{(2,U)}={1\over 2}\left[(\tr\;\Phi)^2+ \tr\,(\Phi^2)\right]
  \qquad \textrm{and} \qquad
  \chi_{(2,U')}={1\over 2}\left[(\tr\;\Phi)^2-\tr\,(\Phi^2)\right]
\end{equation}
are respectively the Schur polynomials of degree $n=2$ in the $U$ and
$U'$ representations\footnote{This is just an example to understand
  the structure of the Schur polynomial, and it must be remembered
  that we always work in the case where $n$ is comparable to $N$.}.

%%%%%%%%%%%%%%%%%%%%%%%%%  
\subsection{Matrix model}

In \cite{Berenstein:2004kk} it was pointed out that, in a particular
regime, GGs could be described by a matrix model. Basically, the idea
is to work in the frame where the CFT lives in a space-time with
$\R\times\calS^3$ topology, and then consider configurations homogeneous
in the $\calS^3$ (i.e. after expanding the CFT fields on spherical
harmonics, we keep only the singlet states). Note that, the only
relevant part of the gauge field ${\bf A}$ is the time-component,
that we can always gauge away keeping in mind that the constraints
tell us to consider only $U(N)$-invariant states. Effectively, we have
reduced the system to a quantum mechanics since only time-dependence
is allowed\footnote{This is very similar to the approximation made
  when GGs are discussed in the test brane picture. There, the static
  gauge is implemented and the degrees of freedom corresponding to
  oscillations of the shape of the GG are frozen.}.

The relevant reduced action was considered in \cite{Hashimoto:2000zp}, 
and reads
\begin{equation}
  S={\Omega_3\ell^3\over\gym^2} \int dt\; 
  \tr\left(\dot{\Phi}^*\dot{\Phi}  - \omega^2\Phi^*\Phi \right)\,,
\label{GGact}\end{equation}
where $\omega=1/\ell$ the inverse radius of AdS$_5$ and $\Omega_3$ is
the volume of the unit three sphere. Also, $*$ stands for complex
conjugation. This action corresponds to the case where we have angular
momentum only in the plane defined by $(X^1,X^4)$. The general
action, with angular momentum in all three planes, corresponds to three
copies of the above action, one for each field $\Phi^1$, $\Phi^2$ and
$\Phi^3$.

Following the usual treatment\footnote{In\cite{Corley:2001zk} similar but different definitions are used.}, we define two
independent matrix valued oscillators $A$ and $B$ by
$A={1\over\sqrt2}(\beta\Phi+\hbox{${i\over \beta}$}\Pi^\dagger)$
and
$B={1\over\sqrt2}(\beta\Phi^\dagger+\hbox{${i\over\beta}$}\Pi)$
with $\beta=\sqrt{\Omega_3\ell^2/\gym^2}$, and the conjugate momenta
$\Pi={\delta S\over\delta\dot\Phi}$. The Hamiltonian $H$ and angular
momentum\footnote{These operators $J^I$ correspond to the angular
  momenta on the plane $(X^I,X^{I+3})$ and measure the R-charge of
  the states.} $J$, in terms of $A$ and $B$, become
\begin{equation}
  H=\omega\,\tr\left(A^\dagger A+B^\dagger B\right),\qquad
  J=\tr\left(A^\dagger A-B^\dagger B\right),
\label{hamiltonian}\end{equation}
and the only nontrivial commutation relations are
\begin{equation}
  \left[A^i_j,(A^\dagger)^k_l\right]=\delta_j^k\delta^i_l\,,\qquad
  \left[B^i_j,(B^\dagger)^k_l\right]=\delta_j^k\delta^i_l\,.
\end{equation}
It follows that $A$ and $B$ are lowering operators for the
hamiltonian, and $A^\dagger$ and $B^\dagger$ are raising operators.
As usual, the vacuum state $\ket{0}$ is defined to be annihilated by
any matrix element of $A$ and $B$. Generic states in the Hilbert space
are then obtained from the vacuum by the repeated action of any matrix
elements of $A^\dagger$ and $B^\dagger$.
Moreover, the application of $A^\dagger$ raises the angular momentum
by one, while $B^\dagger$ lowers it by one. Therefore, the matrix
valued oscillators $A$ and $B$ are left and right handed oscillators
respectively, exciting the two linear harmonic oscillators
$X^1$ and $X^4$ with a phase shift of $\pm\pi/2$. Only the chiral
states, obtained by the exclusive action of $A^\dagger$ (or
$B^\dagger$) are supersymmetric, saturating the BPS bound $H=\omega
J$. Here we will concentrate in the supersymmetric states obtained by
the action of $A^\dagger$ alone.

The matrices $A$ and $B$ admit the so-called polar coordinates
decomposition\footnote{See for example \cite{Klebanov:1991qa} for a
  review in matrix models in string theory and for references.}, where
we write $A=\Omega_A^\dagger\hat A\,\Omega_A$,
$B=\Omega_B^\dagger\hat B\,\Omega_B$ with $\hat A$ and $\hat B$
diagonal matrices and $\Omega_A$ and $\Omega_B$ stand for the angular
variables. Note that we still have the $\Z^N$ invariance acting on the
diagonal matrices. Next, we can always diagonalize one of them, say
$A$ (but not $B$ simultaneously), by acting with a similarity
transformation of $U(N)$. This transformation also changes the measure
in the corresponding path integral of the quantum system. This change
can be reabsorbed by a Vandermonde factor that multiplies the vacuum
state. Since this factor is antisymmetric in the elements of
${\hat A}^\dagger$ and the Hamiltonian still has the $\Z^N$ symmetry
on these same elements, all the excited states will be antisymmetric
in ${\hat A}^\dagger$, rendering our model effectively a fermionic
system (again, we will consider only excitations made out of
$A^\dagger$ to have BPS states).

To be more concrete, we first write
${\hat A}^\dagger=\mathrm{diag}(a^\dagger_1,\dots,a^\dagger_N)$, then
the new normalized vacuum (or {\em fermionic vacuum}) is
\begin{equation}
  \ket{f}={1\over\sqrt{f_v}}\prod_{i<j}(a^\dagger_i-a^\dagger_j)\ket{0},
  \qquad\textrm{with}\quad
  f_v=\prod_{k=1}^N k!\,,
\end{equation}
where the operator in front of $\ket{0}$ is the normalized Vandermonde
determinant. Generic BPS states are written as $S(a^\dagger)\ket{f}$,
with $S(a^\dagger)$ symmetric in $a^\dagger_i$. Therefore, the overall
functions acting on $\ket{0}$ are antisymmetric and can always be
written as a Slater determinant of the following form
\begin{equation}
  \ket{\vec{n}}= {\det\pmatrix{
      & ({a^\dagger_1})^{n_1} & ({a^\dagger_1})^{n_2} & \cdots & ({a^\dagger_1})^{n_N} \cr
      & ({a^\dagger_2})^{n_1} & ({a^\dagger_2})^{n_2} & \cdots & ({a^\dagger_2})^{n_N} \cr
      & \vdots & \vdots & \ddots & \vdots \cr
      & ({a^\dagger_N})^{n_1} & ({a^\dagger_N})^{n_2}& \cdots & ({a^\dagger_N})^{n_N} \cr }
  }
  \ket{0},
\end{equation}
where $\vec n=(n_1,\dots,n_N)$ and we have chosen $n_1>n_2>\ldots>n_N\geq 0$.
Note that $\ket{f}$ corresponds to the minimal occupation
configuration, determining the Fermi sea level and defined by $\vec
n_f=(N-1,N-2,\dots,1,0)$.

At this point, a general state can be represented by the occupation
vector $\vec n$, that in turn has the associated $U(N)$ Young diagram
$$
  \btexdraw
  \drawdim cm
%horizontal
   \move(0 .5)\lvec(4 .5)
   \move(0 0)\lvec(4 0)
   \move(0 -.5)\lvec(3 -.5)
   \move(0 -1.5)\lvec(2.2 -1.5)
   \move(0 -2)\lvec(2.2 -2)
   \move(0 -2.5)\lvec(1.5 -2.5)
%vertical
  \move(0 .5)\lvec(0 -2.5)
  \move(.5 -2.5)\lvec(.5  -1.5)
  \move(.5 -.5)\lvec(.5 .5)
  \move(1 -.5)\lvec(1 .5)
  \move(1   -.0)\lvec(1   .5)
  \move(1.5 -.0)\lvec(1.5 .5)
  \move(4 .0)\lvec(4 .5)
  \move(3 -.5)\lvec(3 0)
  \move(2.2 -2.0)\lvec(2.2 -1.5)
  \move(1.5 -2)\lvec(1.5 -2.5)
%characters
  \move(1.8 .25)\fcir f:0 r:.01
  \move(2.0 .25)\fcir f:0 r:.01
  \move(2.2 .25)\fcir f:0 r:.01
  \move(2.4 .25)\fcir f:0 r:.01
  \move(2.6 .25)\fcir f:0 r:.01
  \move(2.8 .25)\fcir f:0 r:.01
  \move(2.8 .25)\fcir f:0 r:.01
  \move(1.2 -.25)\fcir f:0 r:.01
  \move(1,4 -.25)\fcir f:0 r:.01
  \move(1.6 -.25)\fcir f:0 r:.01
  \move(1.8 -.25)\fcir f:0 r:.01
  \move(2 -.25)\fcir f:0 r:.01
  \move(.6 -1.75)\fcir f:0 r:.01
  \move(.8 -1.75)\fcir f:0 r:.01
  \move(1 -1.75)\fcir f:0 r:.01
  \move(.6 -2.25)\fcir f:0 r:.01
  \move(.8 -2.25)\fcir f:0 r:.01
  \move(.7 -.6)\fcir f:0 r:.01
  \move(.7 -.8)\fcir f:0 r:.01
  \move(.7 -1)\fcir f:0 r:.01
  \move(.7 -1.2)\fcir f:0 r:.01
  \move(.7 -1.4)\fcir f:0 r:.01
  \htext(3.5 .1){$n_1$}
  \htext(2.5 -.4){$n_2$}
  \htext(1.2 -1.9){$n_{N-1}$}
  \htext(.9 -2.4){$n_N$}
  \etexdraw
$$
This diagram tells us that we should take the creation operators
$a_i^\dagger$ to the $n_i$-th power and completely antisymmetrized their
action on the vacuum $\ket{0}$.
In particular, we can relabel the states such that we count the
excitations above the Fermi sea level, by defining a new vector $\vec
n'= \vec n -\vec n_f$. Keeping only the non vanishing $n'_i$, we get
smaller Young diagrams (less rows), telling us only about the
excitations {\em above the Fermi sea level}. Note that,
in our conventions, the operator located at the Fermi sea level is
$a_1^\dagger$ while the one at the bottom corresponds to
$a_N^\dagger$. For example, the following Young diagram
$$
  \btexdraw
  \drawdim cm
%horizontal
   \move(0 .5)\lvec(4 .5)
   \move(0 0)\lvec(4 0)
   \move(0 -.5)\lvec(3 -.5)
%vertical
  \move(0 .5)\lvec(0 -.5)
  \move(.5 .5)\lvec(.5  -.5)
  \move(1 .5)\lvec(1 -.5)
  \move(3  .0)\lvec(3   -.5)
  \move(4 .5)\lvec(4 -.0)
%characters
  \move(1.8 .25)\fcir f:0 r:.01
  \move(2.0 .25)\fcir f:0 r:.01
  \move(2.2 .25)\fcir f:0 r:.01
  \move(2.4 .25)\fcir f:0 r:.01
  \move(2.6 .25)\fcir f:0 r:.01
  \move(2.8 .25)\fcir f:0 r:.01
  \move(2.8 .25)\fcir f:0 r:.01
  %\move(1.2 -.25)\fcir f:0 r:.01
  \move(1,4 -.25)\fcir f:0 r:.01
  \move(1.6 -.25)\fcir f:0 r:.01
  \move(1.8 -.25)\fcir f:0 r:.01
  \move(2 -.25)\fcir f:0 r:.01
  \htext(3.5 .05){$n'_1$}
  \htext(2.5 -.45){$n'_2$}
  \etexdraw
$$
corresponds to exciting only the first two oscillators to the $n'_1$
and $n'_2$ energy levels respectively, over the Fermi sea\footnote{
  Therefore, energy and angular momentum (R-charge) will be
  measured from the Fermi sea level $E_f$.}.

As we have seen, this reduced system comes naturally in terms of
fermionic creation and annihilation operators acting on the vacuum. GGs
were defined in (\ref{GG}) in terms of the complex scalars $\Phi$, and
not of the operators $A$ and $B$. Nevertheless, in the BPS sector of
the Hilbert space we are considering, $A^\dagger$ is proportional to
$\Phi^\dagger$ up to normalization factors (just note that $B$
annihilates any state that we have considered, and then use the
definition of $A^\dagger$ and $B$ to obtain the above
result). Therefore, in this framework, we can describe equivalently
GGs by considering Schur polynomials in $A^\dagger$.

Fortunately, Schur polynomials have already been studied in depth, and
many of their properties are under control. In particular, Schur
polynomials have a very convenient decomposition in terms of Slater
determinants (see for example \cite{libro}) given by
\begin{equation}
  \chi^\dagger_{n,R}=
  \frac{\det(({a^\dagger_j})^{\hat{n}_i+N-i})}
  {\prod_{k<l}\left({a^\dagger}_k-{a^\dagger}_l\right)}\,,
\end{equation}
where the term in the numerator is the determinant of a $N\times N$
matrix labeled by $(i,j)$, ${\hat n}_i$ is the partition of $n$
(i.e. $n=\sum {\hat n_i}$) defined by the specific representation
$R$ of the symmetric group $S_n$ used (${\hat n}_i$ is the number of
boxes in the $i$-th row of the Young tableau associated to the
representation $R$ of $S_n$). In particular, if we are considering GGs
in AdS$_5$, we are instructed to use the $U$ (or totally symmetric)
representation, that has the following partition
\begin{equation}
\hat n_i=n\delta_{i,1}\qquad i=1,\ldots,N,
\end{equation}
but if we consider GG in $\calS^5$, we use the $U'$ representation, with partition
\begin{equation}
\hat n_i=1\qquad i=1,\ldots,N.
\end{equation}
Therefore, a normalized operator representing a GG in AdS acting on the
fermionic vacuum $\ket{f}$ gives
\begin{equation}
\chi^\dagger_{(n,U)}\ket{f}= {1\over\sqrt{f_{ns}f_v}}\;\,{\det \pmatrix{
& (a^\dagger_1)^{n+N-1} & (a^\dagger_1)^{N-2} & \cdots & (a^\dagger_1)^0 \cr
& (a^\dagger_2)^{n+N-1} & (a^\dagger_2)^{N-2} & \cdots & (a^\dagger_2)^0 \cr
& \vdots                & \vdots              & \ddots & \vdots          \cr
& (a^\dagger_N)^{n+N-1} & (a^\dagger_N)^{N-2} & \cdots & (a^\dagger_N)^0 \cr }
}
\ket{0},
\end{equation}
where $f_{ns}=(n+N-1)!/(N-1)!$. Note that, only the first column has
an exponent different from the assigned value characteristic of the
Fermi sea. On the other hand, a normalized GG in $\calS^5$ is given by the
expression
\begin{eqnarray}
&&\!\!\!\!\!\displaystyle\chi^\dagger_{(n,U')}\ket{f}= {1\over \sqrt{f_{na}f_v}}\times
\\
\nonumber\\
&&\qquad\displaystyle\times\,{\det \pmatrix{
    & (a^\dagger_1)^N       & (a^\dagger_1)^{N-1}   & \cdots &
      (a^\dagger_1)^{N-n+1} & (a^\dagger_1)^{N-n-1} & \cdots &
      (a^\dagger_1)^0                                            \cr
    & (a^\dagger_2)^N       & (a^\dagger_2)^{N-1}   & \cdots &
      (a^\dagger_2)^{N-n+1} & (a^\dagger_2)^{N-n-1} & \dots  &
      (a^\dagger_2)^0 \cr
    & \vdots                & \vdots                & \ddots &
      \vdots                & \vdots                & \ddots &
      \vdots                                                     \cr
    & (a^\dagger_N)^N       & (a^\dagger_N)^{N-1}   & \cdots &
      (a^\dagger_N)^{N-n+1} & (a^\dagger_N)^{N-n-1} & \cdots &
      (a^\dagger_N)^0                                            \cr
    }
  }\ket{0},\nonumber
\end{eqnarray}
where $f_{na}=(N-n+1)!/(N-1)!$. Note this time, that there is a
jump in the value of the exponent in the column number $n+1$. Also,
the above states form an orthonormal basis for GGs in AdS, i.e.
$\bra{f}\chi^{\left.\right.}_{(n,U)}{\chi^\dagger}_{(m,U)}\ket{f}=\delta_{nm}$,
and similarly for GGs in $\calS^5$.
Using the rules given before (regarding how to associate a $U(N)$
Young diagram to a given state), both GGs have very simple $U(N)$
diagrams, given respectively as
$$
  \btexdraw
  \drawdim cm
%U
  \move(0 -.0)\lvec(3 -.0)\lvec(3 .5)\lvec(0 .5)\lvec(0 -.0)
  \move(0 0)\lvec(3 0)
  \move( .5 -.0)\lvec(.5  .5)
  \move(1   -.0)\lvec(1   .5)
  \move(1.5 -.0)\lvec(1.5 .5)
  \move(2.5 -.0)\lvec(2.5 .5)
  \move(1.8 .25)\fcir f:0 r:.01
  \move(2.0 .25)\fcir f:0 r:.01
  \move(2.2 .25)\fcir f:0 r:.01
  \htext(.5 -1){GG in AdS$_5$}
%U'
 \move(5 .5)\lvec(5 -2.5)\lvec(5.5 -2.5)\lvec(5.5 .5)\lvec(5 .5)
 \move(5 .0)\lvec(5.5 0)
 \move(5 -.5)\lvec(5.5 -.5)
 \move(5 -2)\lvec(5.5 -2)
 \move(5.25 -.75)\fcir f:0 r:.01
 \move(5.25 -.95)\fcir f:0 r:.01
 \move(5.25 -1.15)\fcir f:0 r:.01
 \htext(6.3 -1){$GG$ in ${\cal S}^5$}
 \etexdraw
$$
where the first diagram is telling us that GGs in AdS$_5$ correspond
to exciting the operator $a^\dagger_1$ (up to the $\Z_N$ action) high
above the Fermi sea level with multiplicity $n$. Also, it is important
to see that, due to the form of the diagram, there is no bound on how
much we can excite this operator, a characteristic feature of GGs in
AdS$_5$. Instead, the behavior of the second diagram is completely
different, and indeed it has a maximum value for $n$ given by
$n=N$. This value is exactly the depth of the Fermi sea, and this time
the interpretation of the diagram is that a GG in $\calS^5$ corresponds to
uplift the Fermi sea level by one unit, creating a hole deep down into
it \cite{Berenstein:2004kk}.

\subsection{Quasi-classical states}

In the above framework, it is useful to remember that each
$a^\dagger_i$ is related to the position of one of the corresponding
D3-branes that form the $U(N)$ theory. Therefore, GGs in AdS can be
understood as the result of separating one D3-brane from the stack of
branes that produces the AdS$_5\times\calS^5$ geometry. This point of
view was already foreseen in \cite{Hashimoto:2000zp}, where a
semi-classical treatment of the action (\ref{GGact}) was
used. Actually, this semi-classical picture seems to capture a lot of
the physics, giving the correct BPS bound, energy and R-charges of the
system, suggesting that we are really facing a quantum system that
somehow behaves classically. This type of phenomenon has already
occurred in the AdS/CFT duality, when large quantum numbers are
involved (see for example \cite{Tseytlin:2003ii}).

To analyse this behavior in the matrix model under consideration, we
shall use {\em coherent states}, defined by the requirement that they
minimize the uncertainty principle. Asking that the quantum uncertainty 
in the energy is much smaller than the measure of the energy itself,
we obtain {\em quasi-classical states}. This further condition
corresponds to consider a large R-charge limit, with the energy well
approximated by the classical value. More precisely, in
appendix~\ref{coherent}, we show that a quasi-classical state
$\ket{\alpha}$ can be defined by
\begin{equation}
\tr(\hat A)\ket{\alpha}=\sqrt k\,\alpha\ket{\alpha}, \qquad 
{\Delta E \over E}\ll 1\,,
\end{equation}
where $E=\bra{\alpha}H\ket{\alpha}-E_f$ is the expectation value of
the energy above the Fermi surface and $\Delta
E^2=\bra{\alpha}H^2\ket{\alpha}-\bra{\alpha}H\ket{\alpha}^2$ measures
its uncertainity. The integer $k\leq N$ defines the number of
D3-branes which are separated from the initial stack of $N$ D3-branes
to build the GGs, if the classical energy is of the form
$E_{classical}=\omega\alpha^*\alpha$.

Indeed, we have observed that quasi-classical states of GGs in AdS$_5$
can be constructed, providing a link between the classical solution of
\cite{Hashimoto:2000zp} and our quantum matrix model. Here we show the
resulting states and the main properties, but a detailed discussion can
be found in appendix \ref{coherent}.

In particular, this set of equations can be solved in the case $k=1$,
describing a single GG in AdS. The resulting coherent state is defined
by (\ref{ggads}),
\begin{equation}
  \ket{\alpha}={1\over \sqrt{f_\alpha}}\sum_{n=0}^{\infty}
  {{\alpha^n\over \sqrt{(N+n-1)!}}}\,\chi^\dagger_{(n,U)}\ket{f},
  \quad\textrm{with}\quad
  f_\alpha=\sum_{n=0}^{\infty}{{(\alpha^*\alpha)^n\over (N+n-1)!}}\,.
  \label{cs}
\end{equation}
This state is explicitly constructed as a superposition of Schur
polynomials in the $U$ representation, which represent a GG in
AdS. Hence, as stated, $\ket\al$ is a coherent GG in AdS. 
As we will show in the next section, this coherent state corresponds
to have a quasi-classical state where only one creation operator
$a_i^\dagger$ (i.e. only one brane) has been excited above
its ground state, reproducing exactly the classical solution of
\cite{Hashimoto:2000zp}.

The second coherent state we achieved to construct is for $k=N$. In
this case, all the $N$ D3-branes are excited with the same
angular momentum, and are described in the CFT by the following state,
\begin{equation}
  \ket {\Omega}={1\over\sqrt{f_N}}
  \exp\left({\al\over\sqrt N}\,\tr\;\hat A^\dagger\right)\ket f, 
  \quad\mathrm{with}\quad f_N=\exp\left(-{\alpha^*\alpha}\right).
\end{equation}

Finally, we showed that it is not possible to construct a coherent
state describing a single GG in $\calS^5$ or, in other words, as
linear combination of Schur polynomials in the $U'$ representation
acting on the Fermi vacuum. This means that such GGs are quantum in
nature, and appear delocalized in space-time from the ten dimensional
point of  view. In particular, their energy is not well defined
classically, and hence the supergravity approximation will fail in the
bulk.

%%%%%%%%%%%%%%%%%%%%%%%%%%%%%%%%%%%%%%%%%%%%%%%%%%%%%%%%%%%%%%%%%%%%%%%%%%%%%
%%%%%%%%%%%%%%%%%%%%%%%%%%%%%%%%%%%%%%%%%%%%%%%%%%%%%%%%%%%%%%%%%%%%%%%%%%%%%%

\section{Back reaction of giant gravitons}
\label{AdS}

In the previous section we described GGs from the point of view of the
CFT theory dual to AdS$_5\times\calS^5$ space-time, in particular we
have obtained quasi-classical states corresponding to GGs in AdS by
means of a coherent superposition of quantum states. On the
gravitational side of the correspondence, this limit can be
interpreted as the localization of the branes in spacetime, and a
description in terms of supergravity fields makes sense in the
bulk. Since these extended objects act as sources for both the
gravitational field and the Ramond-Ramond five-form, their presence
should deform the AdS$_5\times\calS^5$ space-time on which they
live. The purpose of this section is to deduce the form of this back
reaction.

To this end, we evaluate the expectation value of the stress energy
tensor of the CFT theory. Then, using the AdS/CFT correspondence, we
translate it into the boundary stress energy tensor of the
supergravity solution sourced by these GGs, and reconstruct the bulk
fields. Finally, we discuss the validity of this solution and examine
the case of quasi-classical states.

\subsection{General analysis}

First of all, on the CFT side, the configuration we are considering
does not involve gauge fields, as the D3-branes are not
sources for the dilaton and the axion. Moreover, the energy of the GG
does not depend on $\gym$ and therefore the commutator term of the SYM
should not play any role. The relevant part of the CFT action on the
boundary $\R\times\calS^3$ is,
\begin{equation}
  S=-\frac1{2\gym^2}\int d^4x\sqrt{-h}\sum_{m=1}^6
  \tr\left(\p_a X^m\p^a X^m+\frac1{\ell^2}
    (X^m)^2\right)
\end{equation}
where the $1/\ell^2$ term is due to the coupling with the background
curvature, imposed by the conformal invariance. When restricted to
homogeneous configurations this action reduces to three copies of the
model (\ref{GGact}) studied in the previous section.
The associated stress energy tensor $T_{ab}$ is given by
\begin{eqnarray}
  T_{ab}=\frac2{3\gym^2}\sum_{m=1}^{6}
  \tr\left[\nabla_a X^m\nabla_b X^m
    -\frac12X^m\nabla_a\nabla_b X^m +\right.
  \qquad\qquad\qquad\nonumber\\
  -\left.\frac1{4}h_{ab}\left(\nabla_c X^m\nabla^c X^m-2X^m\Box X^m\right)
      +\frac14G_{ab}X^mX^m\right]\,
\end{eqnarray}
where $G_{ab}$ is the Einstein tensor of the background
manifold $\R\times\calS^3$, and $h_{ab}$ is its metric. If we
introduce the unit time vector
$v^a=(1,0,0,0)$, this tensor reads
\begin{equation}
  G_{ab}=-\frac1{\ell^2}\left(h_{ab}-2v_av_b\right)\,.
\end{equation}
Then, assuming that the scalar fields do not depend on the coordinates
on the three-sphere and using the equations of motion, we obtain the
stress tensor
\begin{equation}
  T_{ab}=\frac2{3\gym^2}\sum_{m=1}^6\tr\left(\dot X^m\dot X^m
    +\frac1{\ell^2}X^mX^m\right)\Theta_{ab}\,,
\end{equation}
where we have defined
\begin{equation}
  \Theta_{ab}\equiv v_av_b+\frac14h_{ab}\,.
\label{Theta}\end{equation}
This stress tensor comes in a perfect fluid form and in accordance
with the conformal invariance of the theory, it is traceless.
At this point, it is important to notice that $T_{\mu\nu}$ is
proportional to the hamiltonian (\ref{hamiltonian}) of the reduced
theory. Hence, on the quantum theory the expectation value of the
stress energy tensor in a general state $\ket\psi$ gives
\begin{equation}
  \left<T_{ab}\right>=\frac{4}{3\Omega_3\ell^3}\bra{\psi}H\ket{\psi}
  \Theta_{ab}\,.
\label{qst}\end{equation}
In the full SYM theory, we have here an additional contribution due
to the Casimir effect, which is however known to match the vacuum
contribution of the supergravity theory. By neglecting this term, we
are simply describing the excitations above the vacuum of the theory.

In general,
\begin{equation}
  H=\frac1\ell\sum_{I=1}^{3}\left[J_I+2\;
    \tr\left( B^\dagger_IB_I\right)\right]\,,
\end{equation}
where $J_I$ are the R-charge operators and the trace counts the total
occupation number of the $B_I$ oscillators.
We are interested in BPS states, where only $A^\dagger_I$ excitations
are present. Therefore this trace term does not contribute to the
total energy of the system and the expectation value of the stress
energy tensor is
\begin{equation}
  \left<T_{ab}\right>_{BPS}=\frac{4J}{3\Omega_3\ell^4}\Theta_{ab}\,,
\label{bpsst}\end{equation}
where $J=\left<\sum_IJ_I\right>$ is the expectation value of the total
R-charge. Note that the total energy of the state is
\begin{equation}
  E=\left<T_{tt}\right>\Omega_3\ell^3=\frac{J}{\ell}\,,
\end{equation}
which is the expected relation for BPS states.

The AdS/CFT conjecture tells us that this expectation
value coincides with the boundary stress tensor of the corresponding
supergravity solution.
Hence, the back reaction of the giant graviton on the background is
given, at least asymptotically, by the supergravity solution which has
the correct charges, preserves the same amount of supersymmetries and
whose boundary stress tensor is given by (\ref{qst}). One such
solution is known: it is
given by the supersymmetric Reissner-Nordstr\"om-AdS black hole
studied in \cite{Behrndt:1998ns}, and associated to a condensate of GG
in $\calS^5$ in \cite{superstars}, where this solution was named {\em
  superstar}.

The static R-charged black holes in AdS$_5$, solutions to the $STU$
model, have a metric of the form \cite{Behrndt:1998jd}
\begin{equation}
  ds^2=-H(r)^{-2/3}f(r)dt^2+H(r)^{1/3}\left(\frac{dr^2}{f(r)}+r^2d\Omega_3^2
    \right),
\label{RNmetric}\end{equation}
where the function $f(r)$ is given by
\begin{equation}
  f(r)=1-\frac m{r^2}+\frac{r^2}{\ell^2}H(r)
\end{equation}
and $H(r)$ is the product of three harmonic functions
\begin{equation}
  H(r)=\prod_{I=1}^{3}H_I(r),\qquad
  H_I=1+\frac{q_I}{r^2}.
\end{equation}
Here, $m$ is the non-extremality parameter (determining the mass of the
solution) and the $q_I$ are the R-charges in the $(I,I+3)$ planes of
the five sphere. We parameterize the two scalars of the theory by
three functions $X^I$ constrained by the relation $X^1X^2X^3=1$. These
functions, and the three $U(1)$ gauge fields $\A^I_\mu$, read respectively
\begin{equation}
  X^I=H^{\frac13}H_I^{-1},\qquad
  \A^I=\sqrt{1+\frac{m}{q_{I}}}\left(1-H_I^{-1}\right)\,\dd t.
\end{equation}
In the $m=0$ limit, this solution preserves half of the
supersymmetries  \cite{Behrndt:1998ns}, and a naked singularity develops.
The associated boundary stress energy tensor is computed in
appendix~\ref{stresstensor}, and assumes the form
\begin{equation}
  \hat T_{\mu\nu}=\frac1{6\pi G_5\ell^3}\left(Q+\frac{3\ell^2}8\right)
  \Theta_{\mu\nu}\,,
\end{equation}
where $Q$ is the total R-charge of the system ($Q=\sum_{I=1}^3q_I$).
Here, $G_5$ is the five dimensional Newton's constant, related to the
units of flux $N$ of the five form and to the radius of AdS by
\begin{equation}
  16\pi G_5=\frac{4\Omega_3\ell^3}{N^2}\,.
\end{equation}
Expressed in terms of the CFT constants, the boundary stress tensor becomes
\begin{equation}
  \hat
  T_{\mu\nu}=\frac{2N^2Q}{3\Omega_3\ell^6}\Theta_{\mu\nu}\,,
\end{equation}
where we have dropped the Casimir contribution to the stress energy,
since we are interested to the stress energy of the excitations
above the vacuum.
This result matches perfectly the stress energy tensor (\ref{bpsst})
obtained from the matrix model if we take the R-charges of the CFT state
and those of the supergravity solution to be linked by the relation
\begin{equation}
  \bra{\psi}J_I\ket{\psi}=\frac{N^2}{2\ell^2}q_I\,.
\label{charges}\end{equation}
Note that the charges and boundary stress tensor match also in the
non-extremal case, with the same relation (\ref{charges}) between the
charges, if the total $B_I$ occupation number matches the
non-extremality parameter $m$ of the solution
\begin{equation}
  \bra\psi\sum_{I=1}^3\tr\;B^\dagger_IB_I\ket\psi
  =\frac{3N^2}{8\ell^2} m\,.
\end{equation}
In this case, the supergravity solution would describe a genuine black hole
in AdS$_5$, but the validity of the matrix model description away from
the supersymmetric state is not obvious and will be analyzed elsewhere
\cite{cs}.

\subsection{Classical limit and supergravity description in ten dimensions}

In the general analysis, we have been able to match the expectation
value of the reduced CFT stress tensor with the boundary stress
tensor of a family of supergravity solutions. Note that, in the
matrix model, a generic quantum state $\ket\psi$ corresponds to
excitations of the scalar fields $X^m$ that typically describe a
delocalized system of D3-branes.
In particular, this is also true for the GG states constructed using
single Schur polynomials. On the other hand, we can consider {\em
  quasi-classical} states, representing {\em localized GGs}, that
certainly can be used as sources of supergravity.

For instance, we found in appendix~\ref{coherent} that GGs in
AdS can be excited coherently. In the quasi-classical limit, these
states allow to understand GGs in AdS as set of one or more
D3-branes, separated from the initial stack of $N$ D3-branes used to
define the AdS/CFT duality. In particular, we have constructed
explicitly quasi-classical states corresponding to a single GG in
AdS and the state where all N D3-branes are excited simultaneously. In
contrast, we showed that no analogous coherent states can be defined
for GGs in $\calS^5$, and consequently they have no supergravity
description.

To be more specific, the state corresponding to a single
quasi-classical GG in AdS is given by the coherent state $\ket\alpha$
defined by equation~(\ref{cs}), with large $|\al|$~\footnote{It is
  well known that GGs are well defined as long as their R-charge goes
  at least like $\sqrt N$ \cite{Balasubramanian:2001nh}. We have
  checked that our quasi-classical states have a well defined
  classical limit even for the smaller GGs (see
  appendix~\ref{coherent}).}. In particular, we can take the state with
initial conditions
$\alpha_{10}=\sqrt{\frac{\gym^2J_I}{\Omega_3\ell^2}}$ and
$\alpha_{i0}=0$ for the other $i\geq2$. Then, the scalar
field expectation values behave as a classical solution of the theory,
and take the values
\begin{equation}
  \Phi^1=\sqrt{\frac{\gym^2J_I}{\Omega_3\ell^2}}\hat\eta e^{it/\ell},
\label{classphi}\end{equation}
where $\hat\eta=\mathrm{diag}(1,0,\ldots,0)$. Factoring out the $U(1)$
factor of the $U(N)$ gauge group, corresponding to the motion of the
center of mass of the $N$ D3-branes, $\hat\eta$ becomes the traceless
diagonal $N\times N$ matrix
\begin{equation}
  \hat\eta=\sqrt{\frac{N-1}N}\left(
    \begin{array}{cccc}
      \eta &                 &        &                 \\
           & -\frac\eta{N-1} &        &                 \\
           &                 & \ddots &                 \\
           &                 &        & -\frac\eta{N-1}
    \end{array}
  \right)\,.
\label{eta}\end{equation}
The solution given by equations (\ref{classphi}) and (\ref{eta}) is
exactly the classical CFT solution proposed in \cite{Hashimoto:2000zp}
as dual to a giant graviton in AdS$_5$.

The second system corresponds to the quasi-classical state with all
$\al_i$'s having the same initial conditions (see  equation
(\ref{all})). In this case, we have a collective motion of
all D3-branes, and the giant graviton condensate can be interpreted
as a {\em stack of rotating D3-branes}. This solution, explicitly
given in \cite{Cvetic:1999xp}, corresponds to a type IIB supergravity
solution, which is asymptotically flat and consists in a black
$p3$-brane displaced from the center and rotating in the transverse
$\calS^5$ directions.
By carefully, and simultaneously, taking the near-horizon and extremal
limits, one recovers the supersymmetric solution under consideration.

We would like to stress that these quasi-classical GGs in AdS 
cover all the possible ranges of energies that GGs can describe. In
fact, we are able to obtain GGs with energies $E$, satisfying the
inequality $E\geq \sqrt N$. This same inequality is obtained by
demanding consistency in the test brane picture, where in particular,
the equality corresponds to the case where we have GGs of almost
Planck size \cite{Balasubramanian:2001nh}. For smaller energies,
quantum effects become dominant.

As a last point, using the equation that links the radius of a test GG in
AdS with its energy, $r/\ell=\sqrt{E\ell/N}$ \cite{Hashimoto:2000zp},
we get that its thickness goes to zero for large $N$.
Hence, in the supergravity limit, the branes become infinitely thin,
and the description of classical supergravity plus external
Born-Infeld matter is justified. 
Obviously these are just qualitative arguments, since the relation between the
radius and the energy has been obtained in the test brane
approximation. However, when $r\gg\ell_P$ we expect them to hold, and
indeed they prove the result we already stated: {\em quasi-classical
  states of GGs in AdS are localized in space-time and their back
  reaction is well described by the supersymmetric RN-AdS solution in
  the bulk of AdS$_5\times\calS^5$.}

%%%%%%%%%%%%%%%%%%%%%%%%%%%%%%%%%%%%%%%%%%%%%%%%%%%%%%%%%%%%%%%%%%%%%%%%%%%%%
%%%%%%%%%%%%%%%%%%%%%%%%%%%%%%%%%%%%%%%%%%%%%%%%%%%%%%%%%%%%%%%%%%%%%%%%%%%%%%

\section{Summary and discussion}
\label{END}
In this article we have developed a formalism in the $\NN=4$ SYM to deal with
GGs in terms of a matrix model. This formalism only applies after a
few simplifications have been made. Basically we have focused our
attention to the scalar sector only and considered spatially
homogeneous configurations. Also, we have set the string coupling to
zero to obtain free fields. These approximations are justified for the
states we are interested in, because they excite only the scalar
sector, and the homogeneity corresponds to a freezing of the internal
modes of these object. Moreover, since the states under consideration
are BPS, we expect this description to remain valid as we turn on
$g_s$.
In this framework, GG creation operators are identified with
completely symmetric or antisymmetric Schur polynomials acting on the
Fermi sea describing the vacuum. 

Translating the expectation value of the stress energy tensor into the
boundary stress energy tensor of the corresponding supergravity
solution, we were able to identify the back reaction on the AdS
space-time due to the presence of these branes. The response of the
supergravity fields turned out to be the BPS Reissner-Nordstr\"om-AdS
solution, with one or more R-charges turned on. The resulting field
configuration can be trusted in general only in the asymptotic region,
since for a generic quantum state the giant gravitons are
delocalized. Nevertheless, there are quasi-classical states for which
the branes localize in space-time and, as long as the curvature of the
spacetime remains small, the supergravity solution describes correctly
the back reaction.
These quasi-classical solutions are systems composed by one or more
GGs in AdS. However, there is no analogous quasi-classical state for
GGs in $\calS^5$ and therefore the associated back reacted solution
can only be trusted near the boundary.
In computing the boundary stress tensor of the supergravity solution,
we have also shown that the finite counterterm of Liu and Sabra
\cite{Liu:2004it} is needed to eliminate the scheme-dependent part of
the conformal anomaly.

Regarding future avenues of research, we make notice that in the CFT
matrix model, there is the possibility to discuss non-BPS excitations
acting simultaneously with $A^\dagger$'s and $B^\dagger$'s on the
vacuum. Presumably, if we remain in the near BPS regime, we can still
keep control on the system, opening the way to the description of
static near BPS AdS black holes. Also, the recent discovery of a BPS
black hole in AdS opens up the possibility of an exact derivation of
their entropy. As pointed out by Gutowski and Reall
\cite{Gutowski:2004yv}, these black holes could only be sourced by
D3-branes, or GGs. We believe that, the extension of this matrix model
by adding angular momentum in AdS, could provide the microscopic
degrees of freedom of these objects. In particular, GG configurations
carrying the correct quantum numbers (energy, angular momentum and
R-charges) have been described in \cite{Caldarelli:2004yk}. These
issues are object of current research \cite{cs}.

%%%%%%%%%%%%%%%%%%%%%%%%%%%%%%%%%%%%%%%%%%%%%%%%%%%%%%%%%%%%%%%%%%%%%%%%%%%%%%
%%%%%%%%%%%%%%%%%%%%%%%%%%%%%%%%%%%%%%%%%%%%%%%%%%%%%%%%%%%%%%%%%%%%%%%%%%%%%%

\section*{Acknowledgments}
\small
The authors would like to thank D.~Klemm and M.~Raciti for useful
discussions.

This work was partially supported by INFN, MURST and by the European
Commission RTN program HPRN-CT-2000-00131, in which M.~M.~C. and
P.~J.~S. are associated to the University of Torino.
\normalsize

%%%%%%%%%%%%%%%%%%%%%%%%%%%%%%%%%%%%%%%%%%%%%%%%%%%%%%%%%%%%%%%%%%%%%%%%%%%%%%
%%%%%%%%%%%%%%%%%%%%%%%%%%%%%%%%%%%%%%%%%%%%%%%%%%%%%%%%%%%%%%%%%%%%%%%%%%%%%%

\appendix

\section{Coherent and quasi-classical states}
\label{coherent}

%%% UNCERTAINITY PRINCIPLE
The matrices $X^I$ of the CFT describe the positions of the D3-branes
in spacetime, and for a general quantum state we expect them to be
delocalized. Since we are interested in configurations that admit a
supergravity description, we should look for states that behaves
semi-classically\footnote{In this article, we work exclusively in the Heisenberg picture.}.

A standard approach to obtain such states is to use coherent states,
which have the nice property that they saturate the uncertainity
principle.

\subsection{Coherent states}
%\bigskip
%\noindent{\bf Coherent states}
In the matrix model, due to the $U(N)$ invariance, single
components of our matrices are not observable. Nevertheless, 
the {\em $N$ eigenvalues} of $X^m$ are gauge invariant, 
and describe the positions of the D3-branes (here, the residual $\Z_N$
symmetry corresponds to the fact that the $N$ branes cannot be
distinguished among them).

A particularly interesting gauge-invariant observable is given by the
center of mass of the branes, defined by
\begin{equation}
  Y^m=\frac1N\;\tr\;X^m.
\end{equation}
In particular, this observable carries all the physical information,
for the simple case where a group of $n$ branes are displaced all together
out of the original stack of $N$ D-branes.
In this case, we have that for $X^m$, $n$ eigenvalues are equal to
$\lambda$ and $N-n$ are equal to zero.

The conjugate operators are then given by the total momenta
\begin{equation}
  P_m=\tr\;\pi_m,\qquad\pi_m=\frac{\p\LL}{\p\dot X^m},
\end{equation}
and satisfy the canonical commutation relations
$[Y^m,P_n]=\delta^m_n$. Using standard arguments, one observes that
$\|\left(Y^m-Y^m_0+i\lambda (P_m-P_{0m})\right)\ket\psi\|^2\geq0$ must hold for
any $\lambda$ and deduces that the uncertainty on $Y^m$ and $P_m$ is
bounded by
\begin{equation}
  \Delta Y^m\Delta P_m\geq\frac12\,,
\end{equation}
in units where $\hbar=1$. To achieve minimal uncertainty, we have to
consider states which saturate this inequality and therefore satisfy
$\left(Y^m+i\lambda P_m\right)\ket{\tilde\al}=\tilde\alpha^m\ket{\tilde\al}$
for some $\lambda$. Here, we have labeled the state with a set of
complex numbers, determined by the initial conditions
$\tilde\al^m_0=Y^m_0+i\lambda P_{0m}$.
In addition, we request that $Y^m+i\la P_m$ annihilates the
vacuum of the theory. This condition fixes $\lambda=1/N\be^2$. The
resulting states are called {\em coherent states}, and their defining
relation is given by
\begin{equation}
  \left(\beta NY^m+\frac i\beta P_m\right)\ket\al=\al^m\ket\al.
\label{csdef}\end{equation}

%%% BPS STATES
Let us consider now BPS states. These are defined by the relation
$B^I\ket\psi=0$, which reads in term of the original variables
\begin{equation}
  \left(\beta X^I+\frac i\beta\pi_I\right)\ket\psi
  =i\left(\beta X^{I+3}+\frac i\beta\pi_{I+3}\right)\ket\psi.
\end{equation}
In other words, there is must be a phase shift of $\pi/2$ between
the oscillators $X^I$ and $X^{I+3}$. From this relation, it follows that
\begin{equation}
  \tr\;A^I\ket\psi=\left(\beta NY^I+\frac i\beta P_I\right)\ket\psi
  =i\left(\beta NY^{I+3}+\frac i\beta P_{I+3}\right)\ket\psi.
\end{equation}
Therefore, a BPS coherent state $\ket\al$ in $Y^m$, defined by
(\ref{csdef}) with $\al^I=i\al^{I+3}$,
can be equivalently characterized by the relation
\begin{equation}
  \tr\;A^I\ket\al=\al^I\ket\al,\qquad \al^I\in\C,\qquad I=1\ldots3.
\label{cdef}\end{equation}
We stress the fact that we have to construct $U(N)$-invariant states,
and therefore we cannot simply require the coherence in each element
of the matrices $X^m$. The definition (\ref{cdef}) is however
gauge-invariant and defines a minimal uncertainty state, as required.
Also it is nontrivial to find solutions to this equation, 
since we are working in the case where $A^1=\hat A^1$ is 
diagonal, and  the fermionic Hilbert space is obtained by acting 
with symmetric operators on the Fermi vacuum. Moreover, since we are
mainly interested in condensates of D3-branes, the resulting coherent
states should verify the appropriated symmetry properties.

%%% A STACK OF ROTATING D3-BRANES
\subsection{Particular solutions to the coherent state equation}
%\bigskip
%\noindent {\bf Particular solutions to the coherent state equation}\\
The simplest way to solve equation (\ref{cdef}) is to act with an
exponential of the operator $(\hat A^1)^\dagger$ on the vacuum of the theory,
which is in our case the Fermi vacuum $\ket f$. Taking into account
the phase difference for BPS states, this is equivalent to use an
exponential of $\tr\;(A^1)^\dagger$. Indeed, if we define
\begin{equation}
\label{all}
  \ket {\Omega}={1\over\sqrt{f_N}}\exp\left({\al\over\sqrt N}\tr\;(\hat A^1)^\dagger\right)\ket f, \quad\hbox{with}\quad f_N=\exp\left(-{\alpha^*\alpha}\right)\,,
\end{equation}
we easily obtain using the commutation relations
$[\tr\;\hat A^I,\tr\;\hat A_J^\dagger]=N\delta^I_J$ that indeed this
is a coherent state satisfying $\tr\;\hat A^1\ket {\Omega}=\sqrt N
\al\ket {\Omega}$. Note that we have used a different normalization
for $\alpha$, without loosing generality (see equation (\ref{cdef})),
that will be useful later. Moreover, this state is gauge-invariant, as
it should be, since it corresponds to a symmetric operator acting on
the Fermi vacuum.

%%% JUST ONE COHERENT GIANT GRAVITON IN ADS

In general, a state satisfying (\ref{cdef}) will be coherent in $Y^I$,
however its composition in terms of branes and open strings
excitations is not clear, and may be quite complicated due to the
fermionic structure of the theory.
To have a clear example of quasi-classical GG state, we will now build
a coherent state $\ket{\al}$ describing the excitation of a single GG
in AdS. To have this interpretation, the state must consist of a linear
combination of Schur polynomials $\chi^\dagger_{n,U}$ acting on the
Fermi vacuum. Therefore, we have to find complex coefficients $\ga_n$ such
that the following two equations hold,
\begin{equation}
  \tr\;\hat A^1\ket{\alpha}=\alpha\ket{\alpha},\qquad
  \ket{\alpha}=\sum_{n=1}^\infty \gamma_n \chi^\dagger_{n,\,U}\ket{f}.
\label{tra}\end{equation}
A key relation to solve equation (\ref{tra}), and to obtain an
explicit form of the coherent state $\ket{\alpha}$ is
\begin{equation}
  \tr(\hat A_1)\;\chi^\dagger_{(n,U)}\ket{f}
  =\sqrt{(n+N-1)}\;\chi^\dagger_{(n-1,U)}\ket{f}\;.
\label{key}\end{equation}
To prove this formula, one has to rewrite the ket
$\chi^\dagger_{(n-1,U)}\ket{f}$ in terms of its occupation numbers as
$\ket{n+N-1,N-2,\ldots,1,0}_A$, where the subscript $A$ corresponds to
antisymmetrization. Then, every time that one of the $a_i$ hits a term
of the expansion in the antisymmetrization of the above ket, we get a
term that already existed on the expansion (unless it contains the factor
$(a^\dagger)_i^{n+N-1}$) that, due to the antisymmetrization, will give
zero. Therefore, only the terms with exponent $n+N-1$ will
contribute, producing this number as an overall factor. These
surviving terms have one less operator $a^\dagger_i$, since we have
acted on them with $a_i$, and hence produce a ket of the form
$\ket{n+N-2,N-2,\ldots,1,0}_A$ i.e. a Schur polynomial of degree
$n-1$, $\chi^\dagger_{(n-1,\,U)}$ acting on the Fermi sea. The form of
the final overall value $\sqrt{n+N-1}$ is the result of the ratio of
the difference in the normalization between $\chi^\dagger_{(n,\,U)}$
and $\chi^\dagger_{(n-1,\,U)}$.

At this point, to obtain the final form of the coherent state, we just
have to use the equation (\ref{key}) into equation (\ref{tra}),
getting a recursive series in the $\alpha_n$. Then, the overall
undetermined constant is fixed by the usual normalization
requirement. The explicit form of the state is given by
\begin{equation}
  \label{ggads}
  \ket{\alpha}={1\over\sqrt f_\alpha}
  \sum_{n=0}^{\infty}{{\alpha^n\over\sqrt{(N+n-1)!}}}
  \,\chi^\dagger_{(n,U)}\ket{f},\quad\mathrm{with}\quad
  f_\alpha=\sum_{n=0}^{\infty}{{(\alpha^*\alpha)^n\over (N+n-1)!}}\;,
\end{equation}
where the normalization factor $f_\alpha$ can also be written as,
\begin{equation}
f_\alpha={1\over (\alpha^*\alpha)^{N-1}}
\left(e^{\alpha^*\alpha}-\sum_{n=0}^{N-2}{{(\alpha^*\alpha)^n\over n!}}
\right)\,.
\end{equation}

%%% NO COHERENT GIANTS IN S5

The above procedure can not be implemented exactly for GGs in
$\calS^5$. Basically, the problem is related to the fact that we only
count with a finite number of such independent states, since there are
only $N$ independent Schur polynomials in the $U'$ representation. To
see how the obstruction arises, we first try to solve the analogous
equation of (\ref{tra}),
\bea
\label{tra2}
&&\tr \hat A_1\ket{\beta}=\beta\ket{\beta}\,, \\
&&\ket{\beta}=\sum_{n=1}^N \rho_n \chi^\dagger_{(n,\,U')}\ket{f}\,.
\eea
These equations can be solved by means of the following relation among
normalized GGs in $\calS^5$,
\begin{equation}
  \tr(\hat A_1)\; \chi^\dagger_{(n,U')}\ket{f}=\sqrt{N-n+1}\;
  \chi^\dagger_{(n-1,U')}\ket{f}\;.
\end{equation}
which in turns, is simple to proof with reasoning parallel to the one
used to proof (\ref{key}). After substituting in (\ref{tra2}), we get
the relations
\bea
&&\rho_n={(N-n)!\over N!}(\beta)^n\rho_0\qquad\textrm{for $n=0,\ldots,N$,}
\\
\label{cero}
&&\rho_N=0\,,
\eea
that have as only solution $\rho_n=0$ for any $n$. {\it Therefore it is
  not possible to construct a coherent state made as a superposition
  of GGs in $\calS^5$}.

Nevertheless, we point out that since we work within the AdS/CFT
correspondence, $N$ is considered to be large. We can therefore think
on a combination of GGs in $\calS^5$ that almost satisfy our
constraints, defining an approximated coherent state. For example, we
could arbitrary forget about equation (\ref{cero}), keeping all the
others $\rho_n$. This is certainly an option, but the resulting ket
will spread out in time for any finite $N$, and the classical limit
will loose the very important requirement of localization. The
D3-branes will delocalized after a given scale of time, and from the
ten-dimensional point of view, no supergravity solution will describe
the interior of the corresponding spacetime.

%%% USE OF HAMILTONIAN

%\bigskip
%\noindent {\bf Quasi-classical states}\\
\subsection{Quasi-classical states}
Previously, we have considered the definition of coherent states in
our model.  Nevertheless, we can always impose the extra condition
that the quantum uncertainty in the measurement of the energy $\Delta E$,
is much smaller than the expectation value of the energy $E$. 
In this way, our coherent states reproduce more closely the 
behavior of the classical observables. In this work, 
we will call quasi-classical states, to any coherent states that 
satisfy this extra condition. 

We define our quasi-classical state by means of the
following equations:
\begin{equation}
\tr(\hat A)\ket{\alpha}=\sqrt k\,\alpha\ket{\alpha}, \qquad 
{\Delta E \over E}\ll 1\,,
\label{tran}\end{equation}
where $E=\bra{\alpha}H\ket{\alpha}-E_f$ is the expectation value of
the energy above the Fermi surface and $\Delta
E^2=\bra{\alpha}H^2\ket{\alpha}-\bra{\alpha}H\ket{\alpha}^2$ measures
its uncertainity.
Note that the first equation, corresponding to the 
coherent state equation, has been modified by a factor of $\sqrt k$,
to accommodate the possibility of describing $k$ excited semi-classical
branes. This will be clarified at the end of the appendix when we
discuss the form of the classical solutions of the theory.

Armed with the above definitions, it is not difficult to see that the
coherent state of equation (\ref{all}) is also a quasi-classical
state for the case $k=N$ and therefore is identified 
with the case were all the D3-branes have a total angular momentum in the 
plane defined by $(X^1,X^4)$. In this case, is not difficult to
calculate that 
\begin{equation}
  E=\om\,|\alpha|^2,
\qquad\mathrm{and}\qquad
{\Delta E \over E}= {1\over |\al|},
\end{equation}
which in the limit of large $|\alpha|$ and large $N$, gives the 
desired result. 

Also, the coherent state of equation (\ref{ggads}) is a quasi-classical
state for the case $k=1$, and therefore it must be identified with the
classical solution of a single GG in AdS of \cite{Hashimoto:2000zp}. 

To verify the above statement, we first obtain the form of the energy,
\begin{equation}
E={\om\over f_\al}\sum_{n=0}^{\infty}
  {{n\left(\alpha^*\alpha\right)^n\over\sqrt{(N+n-1)!}}},
\end{equation}
which can be rewritten as
\begin{equation}
  E=\om|\al|^2-{\om\gamma(N,|\al|^2)\over\gamma(N-1,|\al|^2)}
\end{equation}
using the incomplete gamma function. Then, after some calculations, 
we obtain that
\begin{equation}
  {\Delta E\over E} = \left({|\al|^2\over E^2}-{|\al|^2
      \exp\left(|\al|^2\right)\over
      |\al|^2\,\gamma(N-1,|\al|^2)-\gamma(N,|\al|^2)}\right)^{1/2}\,.
\end{equation}
Let us consider now two different large $|\al|$ limits, that define
quasi-classical regimes. The first one corresponds to take 
$|\al|^2\gg N$, that gives
\begin{equation}
E=\omega |\al|^2 \,+\OO\left({1\over|\al|^2}\right),
\qquad\mathrm{and}\qquad
{\Delta E\over E}={1\over|\al|^2}+\OO\left({1\over|\al|^3}\right).
\end{equation}
A second asymptotic expansion can be obtained by taking
$|\al|^2=N+\sqrt{2N}y$, where $y\in\R$ is bounded and $N$ large. Here
we found that
\begin{equation}
E=\omega\sqrt{2N}\left[y+g(y)\right]+\OO\left(N^0\right),
\end{equation}
\begin{equation}
{\Delta E^2\over E^2}=\frac1{2\left(y+g(y)\right)^{2}}
-\frac{g(y)}{y+g(y)}\,+\,\OO\left({1\over \sqrt N}\right).
\end{equation}
The function $g(y)$ is defined by
\begin{equation}
g(y)={e^{-y^2}\over \sqrt{\pi}\,\erfc(-y)}\,,
\end{equation}
and vanishes very rapidly when $y$ is positive. Hence, the above
expressions can be approximated for $y\gtrsim1$ by
\begin{equation}
E\approx\omega\,\sqrt{2N}\,y\,,
\qquad\mathrm{and}\qquad
{\Delta E \over E}\approx {1\over \sqrt{2}\,\! y}\,.
\end{equation}
Remember that, in the test brane picture, the minimal energy that can
be reached is precisely of order $\sqrt N$ \cite{Balasubramanian:2001nh}.
Therefore, putting together the two limiting cases described above we
cover all the possible energy ranges of a GG in AdS that can be
described in the test brane picture, from the smallest
$E\sim\sqrt{N}/\ell$ GGs to the cosmological ones.

As a last point, we review the form of the classical solution, to
justify the interpretation we have made on the number of classical
GGs in a given quasi-classical state.
In the classical theory, the relevant part of
the hamiltonian is of the form
$H_{classical}=\omega\;\tr\left(\alpha^*\alpha\right)$ with
$\alpha={1\over\sqrt 2}\left(\beta\Phi+\frac i\beta\Pi^*\right)$ and
$\Pi_I={\p\LL\over\p\dot \Phi^I}$. After having diagonalized the
matrix $\alpha$, the classical equations of motion for the diagonal
elements $\alpha_i$ have the general solution
$\alpha_i(t)=\alpha_{i0}e^{-i\omega\,t}$, with energy
$E=\omega\,\sum_{i=1}^N \alpha_{i0}^*\alpha_{i0}$. Here, the coefficients
$\alpha_{i0}$ encode the information of the initial conditions.

Following the interpretation given in \cite{Hashimoto:2000zp}
of the above classical system, consider the case where $k$ oscillators
have the same initial condition $\al_{i0}=\sqrt{k}\al_0$ for $i=1,\dots,k$
and all others are zero.
This system corresponds to have $k$ GGs with the same energy.
If we evaluate the hamiltonian $H$, and the operator $\tr\,\al$,
in this  configuration, we get $H_{classical}=\,\omega\,\alpha_0^*\alpha_0$,
and $\tr\,\alpha= \sqrt{k} \alpha_0e^{-iwt}$. Note that, the above relation
characterizes the classical state of $k$ GGs in AdS. In particular in the
$k=1$ case,
we are in the presence of a single GG, while for $k=N$, we are in the
case where all the branes have the same R-charge. This last two
extreme cases correspond to different identifications of the $U(1)$ in
$U(N)$.

In the quantum system, the hamiltonian is again the sum of $N$
harmonic oscillators, but we have to consider only gauge invariant
observables, antisymmetrized in the $N$ oscillators variables. Equation 
(\ref{cdef}) gives our generalized definition of coherent state. The
operator $\tr\;\hat A^1$ has an equation of motion of the same form as the
single operators $a_i$'s, with general solution
$\tr\hat A^1=(\tr\hat A^1)_{0}e^{-i\omega\,t}$. 
At this point  $(\tr \hat A^1)_0$ is a complex number defining the initial 
condition (note that we have restricted only the total sum of the
initial condition of each operator $a_i$, and therefore many different
combinations of initial conditions in $a_i$ will have the same total
initial condition in $\tr \hat A^1$ in the same manner as for the
classical theory).

Therefore, the use of equation (\ref{tran}) guaranties that the corresponding 
quasi-classical state has the classical interpretation of $k$ displaced branes.

%%%%%%%%%%%%%%%%%%%%%%%%%%%%%%%%%%%%%%%%%%%%%%%%%%%%%%%%%%%%%%%%%%%%%%%%%%%%%%
%%%%%%%%%%%%%%%%%%%%%%%%%%%%%%%%%%%%%%%%%%%%%%%%%%%%%%%%%%%%%%%%%%%%%%%%%%%%%%

\section{Boundary stress tensor}
\label{stresstensor}

In this appendix we compute the boundary stress energy tensor for the
Reissner-Nordstr\"om-AdS$_5$ family of black holes with three
independent charges in the $STU$ model. It is defined by the
quasi-local stress tensor of Brown and York \cite{Brown:1992br}, as
the variation of the gravitational action $S_0$ (including the Gibbons
Hawking surface term to have a well-defined variational principle
\cite{Gibbons:1976ue}) with respect to the metric
$g_{\mu\nu}$ induced on the boundary $\p\MM$ of the manifold,
\begin{equation}
  T^{\mu\nu}=\frac2{\sqrt{-g}}\frac{\delta S_0}{\delta g_{\mu\nu}}\,.
\end{equation}
The action, and the stress tensor, are regularized by taking the
boundary at finite radius $r$. As $\p\MM$ is pushed to infinity, the
stress tensor diverges. In the framework of AdS/CFT it is natural to
renormalize the supergravity action by the addition of local
counterterms on the boundary \cite{Balasubramanian:1999re}. This
method has the advantage of being background independent and
covariant. In AdS$_5$, the needed counterterm has the form
\cite{Balasubramanian:1999re,Emparan:1999pm}
\begin{equation}
  S_{ct}[g_{\mu\nu}]=\frac1{8\pi
  G_5}\int_{\MM}\sqrt{-g}\left[-\frac3{\ell}
  \left(1-\frac{\ell^2}{12}\RR\right)\right]\,,
\end{equation}
where $\RR$ is the curvature scalar of the boundary metric. There is
still the freedom to add {\em finite counterterms}, which are needed
if one wants to recover for example a consistent thermodynamics for
these black holes \cite{klemm}. In \cite{Liu:2004it}, inclusion of
such counterterms was advocated in order to recover the expected ADM
energies and a BPS-like linear relation between energy and
R-charges. This corresponds to a shift in the renormalization
prescription in the dual CFT. 

What we show in this appendix, is that the deep
reason for the introduction of this term is to {\em eliminate the
scheme-dependent part of the Weyl anomaly}. Hence, by requiring a
regularization procedure which respects the full conformal symmetry,
we are able to lift the ambiguity in the definition of the action and
conserved charges of the theory.

For the three-charges Reissner-Nordstr\"om-AdS$_5$ solutions
(\ref{RNmetric}), the quasi-local stress tensor obtained from the
action $S_0+S_{ct}$ reads
\begin{equation}
  8\pi G_5T_{\mu\nu}=\frac4{3\ell r^2}\left(\frac{3\ell^2}8+\frac32m+Q\right)
  \Theta_{\mu\nu}-\frac1{\ell^3r^2}\left(\tilde
    Q-\frac13Q^2\right)g_{\mu\nu}
  +\OO\left(\frac1{r^4}\right)\,,
\label{qlst}\end{equation}
where $Q=\sum_Iq^I$ is the total R-charge and $\tilde
Q=\sum_{I<J}q^Iq^J$, and $\Theta_{\mu\nu}$ is the tensor defined by
(\ref{Theta}). Note that there is a trace anomaly proportional to
$\tilde Q-Q^2/3$, which originates from a choice of renormalization
scheme which does not respect the asymptotic isometry group of
AdS$_5$. From the CFT point of view, the regularization used violates
the scale invariance. Indeed, the general form of the Weyl anomaly
consists of a scheme independent part, which vanishes for the
background under consideration, and a collection of total derivative
terms that can be removed by adding suitable finite counterterms to
the action \cite{Deser:yx,Henningson:1998gx}. Hence, the anomaly we
obtain is trivial, and it is possible to restore the conformal
symmetry by adding to the action precisely the finite counterterm
proposed by Liu and Sabra \cite{Liu:2004it}. For the class of
solutions (\ref{RNmetric}) under consideration, its exact form is
\begin{equation}
  S_{\phi^2}=\frac1{8\pi G_5}
  \int_{\MM}\sqrt{-g}\left(\frac1{2\ell}h_{ij}\phi^i\phi^j\right)\,,
\end{equation}
where $\phi^i$, $i=1,2$ parameterize the two scalars of the $STU$
model, and $h_{ij}$ is the moduli space metric.
The contribution of the finite $\phi^2$ counterterm is easily computed
and is
\begin{equation}
  8\pi G_5T^{\phi^2}_{\mu\nu}=\frac1{\ell^3r^2}\left(\tilde
    Q-\frac13Q^2\right)g_{\mu\nu}
  +\OO\left(\frac1{r^4}\right)\,,
\end{equation}
which is precisely the quantity needed to cancel the trace anomaly
from the quasi-local stress tensor (\ref{qlst}), and eliminate the
nonlinear dependence on the charges. Finally, the complete
quasi-local stress tensor reads,
\begin{equation}
  8\pi G_5T_{\mu\nu}=\frac4{3\ell r^2}\left(\frac{3\ell^2}8+\frac32m+Q\right)
  \Theta_{\mu\nu}+\OO\left(\frac1{r^4}\right)\,.
\end{equation}
To find the boundary stress tensor, we push the boundary $\p\MM$ to
infinity, conformally rescaling the metric in such a way to eliminate
the double pole. Defining the background metric upon which the dual
field theory resides as
\begin{equation}
  h_{\mu\nu}=\lim_{r\rightarrow\infty}\frac{\ell^2}{r^2}g_{\mu\nu}
\end{equation}
we find that the boundary is the Einstein universe with metric
$ds^2=-dt^2+\ell^2d\Omega_3^2$.
Hence, the CFT stress tensor is obtained by rescaling the quasi-local
stress tensor by a factor $r^2/\ell^2$ before taking the boundary at
infinity. As a result, one obtains
\begin{equation}
  8\pi G_5T_{\mu\nu}=\frac4{3\ell^3}\left(\frac{3\ell^2}8+\frac32m+Q\right)
  \Theta_{\mu\nu}\,.
\label{bst}\end{equation}
The AdS/CFT correspondence tells us that (\ref{bst}) coincides with
the quantum expectation value of the CFT stress tensor:
\begin{equation}
  \left<T^{CFT}_{\mu\nu}\right>=T_{\mu\nu}\,.
\end{equation}
The first term in the parenthesis of eqn.~\ref{bst} corresponds to the
non-vanishing vacuum energy due to the Casimir effect of the CFT on
$\R\times\calS^3$, and the term proportional to $3m/2+Q$ is the energy
of the excitation over the vacuum.

%%%%%%%%%%%%%%%%%%%%%%%%%%%%%%%%%%%%%%%%%%%%%%%%%%%%%%%%%%%%%%%%%%%%%%%%%%%%%%
%%%%%%%%%%%%%%%%%%%%%%%%%%%%%%%%%%%%%%%%%%%%%%%%%%%%%%%%%%%%%%%%%%%%%%%%%%%%%%


\begin{thebibliography}{99}


%\cite{Petersen:1999zh}
\bibitem{Petersen:1999zh}
J.~L.~Petersen,
``Introduction to the Maldacena conjecture on AdS/CFT,''
Int.\ J.\ Mod.\ Phys.\ A {\bf 14} (1999) 3597
[arXiv:hep-th/9902131].
%%CITATION = HEP-TH 9902131;%%

%\cite{Aharony:1999ti}
\bibitem{Aharony:1999ti}
O.~Aharony, S.~S.~Gubser, J.~M.~Maldacena, H.~Ooguri and Y.~Oz,
``Large N field theories, string theory and gravity,''
Phys.\ Rept.\  {\bf 323} (2000) 183
[arXiv:hep-th/9905111].
%%CITATION = HEP-TH 9905111;%%

%\cite{D'Hoker:2002aw}
\bibitem{D'Hoker:2002aw}
E.~D'Hoker and D.~Z.~Freedman,
``Supersymmetric gauge theories and the AdS/CFT correspondence,''
arXiv:hep-th/0201253.
%%CITATION = HEP-TH 0201253;%%

%\cite{McGreevy:2000cw}
\bibitem{McGreevy:2000cw}
J.~McGreevy, L.~Susskind and N.~Toumbas,
``Invasion of the giant gravitons from anti-de Sitter space,''
JHEP {\bf 0006} (2000) 008
[arXiv:hep-th/0003075].
%%CITATION = HEP-TH 0003075;%%

%\cite{Caldarelli:2004yk}
\bibitem{Caldarelli:2004yk}
M.~M.~Caldarelli and P.~J.~Silva,
``Multi giant graviton systems, SUSY breaking and CFT,''
JHEP {\bf 0402}, 052 (2004)
[arXiv:hep-th/0401213].
%%CITATION = HEP-TH 0401213;%%

%\cite{Hackett-Jones:2004yi}
\bibitem{Hackett-Jones:2004yi}
E.~J.~Hackett-Jones and D.~J.~Smith,
``Type IIB Killing spinors and calibrations,''
arXiv:hep-th/0405098.
%%CITATION = HEP-TH 0405098;%%

%\cite{nonsusy}
\bibitem{nonsusy}
%\cite{Page:2002xz}
%\bibitem{Page:2002xz}
D.~C.~Page and D.~J.~Smith,
``Giant gravitons in non-supersymmetric backgrounds,''
JHEP {\bf 0207} (2002) 028
[arXiv:hep-th/0204209];\\
%%CITATION = HEP-TH 0204209;%%
%\cite{Hackett-Jones:2002ph}
%\bibitem{Hackett-Jones:2002ph}
E.~J.~Hackett-Jones and D.~J.~Smith,
``Giant gravitons as probes of gauged supergravity,''
JHEP {\bf 0210} (2002) 075
[arXiv:hep-th/0208015].
%%CITATION = HEP-TH 0208015;%%

%\cite{Hashimoto:2000zp}
\bibitem{Hashimoto:2000zp}
A.~Hashimoto, S.~Hirano and N.~Itzhaki,
``Large branes in AdS and their field theory dual,''
JHEP {\bf 0008} (2000) 051
[arXiv:hep-th/0008016];\\
%%CITATION = HEP-TH 0008016;%%

%\cite{Grisaru:2000zn}
\bibitem{Grisaru:2000zn}
M.~T.~Grisaru, R.~C.~Myers and O.~Tafjord,
``SUSY and Goliath,''
JHEP {\bf 0008} (2000) 040
[arXiv:hep-th/0008015].
%%CITATION = HEP-TH 0008015;%%

%\cite{Maldacena:1998bw}
\bibitem{Maldacena:1998bw}
J.~M.~Maldacena and A.~Strominger,
``AdS(3) black holes and a stringy exclusion principle,''
JHEP {\bf 9812} (1998) 005
[arXiv:hep-th/9804085].
%%CITATION = HEP-TH 9804085;%%

%\cite{Balasubramanian:2001nh}
\bibitem{Balasubramanian:2001nh}
V.~Balasubramanian, M.~Berkooz, A.~Naqvi and M.~J.~Strassler,
``Giant gravitons in conformal field theory,''
JHEP {\bf 0204} (2002) 034
[arXiv:hep-th/0107119].
%%CITATION = HEP-TH 0107119;%%

%\cite{Bena:2004qv}
\bibitem{Bena:2004qv}
I.~Bena and D.~Smith,
``Towards the solution to the giant graviton puzzle,''
arXiv:hep-th/0401173.
%%CITATION = HEP-TH 0401173;%%

%\cite{Corley:2001zk}
\bibitem{Corley:2001zk}
S.~Corley, A.~Jevicki and S.~Ramgoolam,
``Exact correlators of giant gravitons from dual N = 4 SYM theory,''
Adv.\ Theor.\ Math.\ Phys.\  {\bf 5} (2002) 809
[arXiv:hep-th/0111222].
%%CITATION = HEP-TH 0111222;%%

%\cite{Berenstein:2004kk}
\bibitem{Berenstein:2004kk}
D.~Berenstein,
``A toy model for the AdS/CFT correspondence,''
arXiv:hep-th/0403110.
%%CITATION = HEP-TH 0403110;%%

%\cite{Behrndt:1998ns}
\bibitem{Behrndt:1998ns}
K.~Behrndt, A.~H.~Chamseddine and W.~A.~Sabra,
``BPS black holes in N = 2 five dimensional AdS supergravity,''
Phys.\ Lett.\ B {\bf 442} (1998) 97
[arXiv:hep-th/9807187].
%%CITATION = HEP-TH 9807187;%%

%\cite{superstars}
\bibitem{superstars}
R.~C.~Myers and O.~Tafjord,
``Superstars and giant gravitons,''
JHEP {\bf 0111} (2001) 009
[arXiv:hep-th/0109127];\\
%%CITATION = HEP-TH 0109127;%%
%\cite{Leblond:2001gn}
%\bibitem{Leblond:2001gn}
F.~Leblond, R.~C.~Myers and D.~C.~Page,
``Superstars and giant gravitons in M-theory,''
JHEP {\bf 0201} (2002) 026
[arXiv:hep-th/0111178].
%%CITATION = HEP-TH 0111178;%%

%\cite{Liu:2004it}
\bibitem{Liu:2004it}
J.~T.~Liu and W.~A.~Sabra,
``Mass in anti-de Sitter spaces,''
arXiv:hep-th/0405171.
%%CITATION = HEP-TH 0405171;%%

\bibitem{odintsov}
%\cite{Nojiri:2002td}
%\bibitem{Nojiri:2002td}
S.~Nojiri and S.~D.~Odintsov,
``Universal features of the holographic duality: Conformal anomaly and brane
gravity trapping from 5D AdS black hole,''
Int.\ J.\ Mod.\ Phys.\ A {\bf 18} (2003) 2001
[arXiv:hep-th/0211023];\\
%%CITATION = HEP-TH 0211023;%%
%\cite{Cvetic:2003zy}
%\bibitem{Cvetic:2003zy}
M.~Cvetic, S.~Nojiri and S.~D.~Odintsov,
``Cosmological anti-deSitter space-times and time-dependent AdS/CFT
correspondence,''
Phys.\ Rev.\ D {\bf 69} (2004) 023513
[arXiv:hep-th/0306031].
%%CITATION = HEP-TH 0306031;%%

%\cite{Klebanov:1991qa}
\bibitem{Klebanov:1991qa}
I.~R.~Klebanov,
``String theory in two-dimensions,''
arXiv:hep-th/9108019.
%%CITATION = HEP-TH 9108019;%%

%\cite{libro}
\bibitem{libro}
W. Fulton and J. Harris,
``Representatin Theory'', Graduated text in Mathematicl physics 129,
Springer-Verlag (1991), J.H. Ewing F.W. Gehring P.R. Halamos (Editors)

%\cite{Tseytlin:2003ii}
\bibitem{Tseytlin:2003ii}
A.~A.~Tseytlin,
``Spinning strings and AdS/CFT duality,''
arXiv:hep-th/0311139.
%%CITATION = HEP-TH 0311139;%%

%\cite{Behrndt:1998jd}
\bibitem{Behrndt:1998jd}
K.~Behrndt, M.~Cvetic and W.~A.~Sabra,
``Non-extreme black holes of five dimensional N = 2 AdS supergravity,''
Nucl.\ Phys.\ B {\bf 553} (1999) 317
[arXiv:hep-th/9810227].
%%CITATION = HEP-TH 9810227;%%

%\cite{cs}
\bibitem{cs}
M.~M.~Caldarelli and P.~J.~Silva,
in preparation.

%\cite{Cvetic:1999xp}
\bibitem{Cvetic:1999xp}
M.~Cvetic {\it et al.},
``Embedding AdS black holes in ten and eleven dimensions,''
Nucl.\ Phys.\ B {\bf 558} (1999) 96
[arXiv:hep-th/9903214].
%%CITATION = HEP-TH 9903214;%%

%\cite{Gutowski:2004yv}
\bibitem{Gutowski:2004yv}
J.~B.~Gutowski and H.~S.~Reall,
``General supersymmetric AdS(5) black holes,''
JHEP {\bf 0404}, 048 (2004)
[arXiv:hep-th/0401129].
%%CITATION = HEP-TH 0401129;%%
%\cite{Gutowski:2004ez}
%\bibitem{Gutowski:2004ez}
\\
J.~B.~Gutowski and H.~S.~Reall,
``Supersymmetric AdS(5) black holes,''
JHEP {\bf 0402}, 006 (2004)
[arXiv:hep-th/0401042].
%%CITATION = HEP-TH 0401042;%%

%\cite{Brown:1992br}
\bibitem{Brown:1992br}
J.~D.~Brown and J.~W.~.~York,
``Quasilocal energy and conserved charges derived from the gravitational
action,''
Phys.\ Rev.\ D {\bf 47} (1993) 1407.
%%CITATION = PHRVA,D47,1407;%%

%\cite{Gibbons:1976ue}
\bibitem{Gibbons:1976ue}
G.~W.~Gibbons and S.~W.~Hawking,
``Action Integrals And Partition Functions In Quantum Gravity,''
Phys.\ Rev.\ D {\bf 15} (1977) 2752.
%%CITATION = PHRVA,D15,2752;%%

%\cite{Balasubramanian:1999re}
\bibitem{Balasubramanian:1999re}
V.~Balasubramanian and P.~Kraus,
``A stress tensor for anti-de Sitter gravity,''
Commun.\ Math.\ Phys.\  {\bf 208} (1999) 413
[arXiv:hep-th/9902121].
%%CITATION = HEP-TH 9902121;%%

%\cite{Emparan:1999pm}
\bibitem{Emparan:1999pm}
R.~Emparan, C.~V.~Johnson and R.~C.~Myers,
``Surface terms as counterterms in the AdS/CFT correspondence,''
Phys.\ Rev.\ D {\bf 60} (1999) 104001
[arXiv:hep-th/9903238].
%%CITATION = HEP-TH 9903238;%%

\bibitem{klemm}
D.~Klemm, private communication.

%\cite{Deser:yx}
\bibitem{Deser:yx}
S.~Deser and A.~Schwimmer,
``Geometric Classification Of Conformal Anomalies In Arbitrary Dimensions,''
Phys.\ Lett.\ B {\bf 309} (1993) 279
[arXiv:hep-th/9302047].
%%CITATION = HEP-TH 9302047;%%

%\cite{Henningson:1998gx}
\bibitem{Henningson:1998gx}
M.~Henningson and K.~Skenderis,
``The holographic Weyl anomaly,''
JHEP {\bf 9807} (1998) 023
[arXiv:hep-th/9806087].
%%CITATION = HEP-TH 9806087;%%

\end{thebibliography}
\end{document}